\documentclass[aps,twocolumn,pra,superscript,floatfix,superscriptaddress,showpacs]{revtex4-1}

\usepackage{xcolor}
\usepackage{epsf}
\usepackage{epsfig}
\usepackage{graphicx}

\usepackage{dcolumn}
\usepackage{braket}
\usepackage{bm}
\usepackage{amsfonts}
\usepackage{amsmath}
\usepackage{amssymb}
\usepackage{color,soul}
\usepackage{wasysym}
\usepackage{mathrsfs}
\usepackage{natbib}
\usepackage{float}
\usepackage{multirow}

\usepackage[colorlinks=true,%
            linkcolor=blue,%
            urlcolor=blue,%
            citecolor=blue,%
            filecolor=blue,%
            bookmarksopen=true,%
            pdfauthor={},%
            pdftitle={BIC},%
            pdfsubject={BIC_Manuscript},%
            pdfpagemode=UseOutlines]{hyperref}

\definecolor{otrogreen}{rgb}{0, 0.7, 0.}
\definecolor{antiquefuchsia}{rgb}{0.57, 0.36, 0.51}
\definecolor{amethyst}{rgb}{0.6, 0.4, 0.8}

\begin{document}

\title{Tunable quantum photonic routing using a coupled giant-atom-like array}
\author{Alexis R. Leg\'{o}n}
\affiliation{Departamento de F\'{\i}sica, Universidad T\'{e}cnica Federico Santa Mar\'{i}a, Casilla 110 V, Valparaiso, Chile}
\author{Mario Miranda}
\affiliation{Departamento de F\'{\i}sica, Universidad T\'{e}cnica Federico Santa Mar\'{i}a, Casilla 110 V, Valparaiso, Chile}
\author{P.\ A.\ Orellana}
\affiliation{Departamento de F\'{\i}sica, Universidad T\'{e}cnica Federico Santa Mar\'{i}a, Casilla 110 V, Valparaiso, Chile}
\

\begin{abstract}
We examine a quantum routing mechanism utilizing a giant-atom-like array coupled to two one-dimensional waveguides. The giant-atom-like array is formed by a one-dimensional array of three-level-systems. In the regime of strong atom-waveguide coupling and weak inter-atomic interactions, this system functions as an efficient and directionally controllable single-photon router. Our analysis shows that the routing behavior is influenced by effective phase accumulation and interference effects, which can be adjusted by varying the number of coupling sites $N$, the photon energy $E$, and the inter-atomic coupling strength $J$. 
Importantly, we identify configurations that enable perfect photon transfer ($100 \%$ efficiency) over a wide range of energies and that provide dynamic control over the output channel. In addition, we investigate how the system responds to changes in its internal parameters, demonstrating the robustness and scalability of routing performance. These findings underscore the potential of this setup for implementation in reconfigurable and integrated quantum photonic networks.
\end{abstract}

\maketitle

\section{Introduction}
Recent advancements in quantum technologies have opened up new possibilities for processing information, exceeding the capabilities of classical systems. In this context, quantum networks have emerged as a key platform for developing scalable technologies in quantum communication, distributed quantum computing, and information processing ~\cite{Reiserer2015,Wei2022,Abane2025}. These networks enable the coherent sharing of quantum states between distant nodes, primarily using photons that travel through specially engineered waveguides or integrated photonic circuits~\cite{Li2022,Roy2017}.

A quantum router is any device that enables the control of the path taken by a single photon as it travels through a quantum network. Over the past decade, a variety of experimental platforms have been explored for implementing quantum routers, including cavity quantum electrodynamics (QED) systems~\cite{Chang2018, Turschmann2019, Sheremet2023}, superconducting circuits~\cite{Chapman2017, Wang2021, Li2024}, and integrated photonic waveguides~\cite{Vetsch2010, Silverstone2016}. Among these options, waveguide quantum electrodynamics (waveguide QED) is notable for its capacity to support strong and controllable interactions between photons and quantum emitters at the level of individual quanta.

In recent years, \textit{giant atoms}—quantum emitters connected to a waveguide at multiple distant points—have generated significant interest. This type of non-local coupling introduces interference effects that do not occur in standard, locally coupled systems. It has paved the way for a range of remarkable quantum optical phenomena, including decoherence-free subspaces, non-Markovian behavior, and bound states embedded in the continuum ~\cite{Calajo2019,Facchi2016,Kockum2018,Guimond2020}. Due to these unique properties, giant atoms are now regarded as promising building blocks for tunable, high-efficiency quantum routers.

Most existing routing schemes that utilize giant-atom platforms still encounter challenges such as limited efficiency, a lack of directional control, and restricted tunability in photon transfer~\cite{Soro2022,Yin2023,Zhang_Yu2023}. In addition to photon transport, giant atoms have been investigated as simulators of open quantum systems. Their non-local couplings enable the emulation of non-Hermitian dynamics and the quantum Zeno crossover, allows to construct scalable digital–analog quantum simulations~\cite{Chen2025}. While some experimental efforts have successfully demonstrated chiral coupling and multi-port routing capabilities~\cite{Wang2021,Li2024,Sollner2015,Zhou2015,Wang2022Chiral}, practical implementations of reconfigurable or robust architectures remain limited~\cite{Soro2022,Zhang_Yu2023,Yin2023,Zheng2024,Wang2025,Luo2024,Noachtar2022,Peng2023,TerradasBrianso2022}. These limitations highlight the need for novel strategies that makes possible interference and non-local photon–atom interactions in engineered quantum media.

Motivated by recent developments, we investigate a new class of systems that operate at the interface of condensed matter physics and photonic metamaterials\cite{Kannan2020,Zanner2022}. Specifically, we focus on arrays of metamaterial resonators coupled to two one-dimensional transmission lines, creating a synthetic waveguide QED platform \cite{Du2022,Yang2022}. In previous studies, it was explored photon dynamics in a system consisting of three-level atoms coupled to two distinct waveguide channels, demonstrating its effectiveness as a single-photon router ~\cite{Huang2018,Ahumada2019,Ahumada2022}. Building on these findings, we aim to extend our approach by using giant atoms and tailored non-local couplings to achieve deterministic and controllable quantum routing\cite{Zheng2024,Wang2025}.

In this work, we explore a quantum routing mechanism that utilizes a coupled array of atoms behaving as a giant atom, connected to two one-dimensional waveguides. We demonstrate that, under conditions of strong coupling between the atoms and the waveguides, along with weak interactions between the atoms, the system can effectively serve as a directionally controllable single-photon router. Our analysis highlights the effectiveness of phase accumulation and destructive interference in influencing routing behavior, thereby aiding in the design of integrated and reconfigurable quantum photonic devices. Notably, we identify configurations that allow for nearly $100\%$ photon transfer across a broad energy range around the center of the band, as well as setups that enable controlled and tunable routing of photons into either waveguide. These findings point to the potential development of advanced, integrated, and reconfigurable quantum photonic devices.

\section{Model}
\label{sec:1}

\begin{figure}[htbp]
\includegraphics[width=1.05\columnwidth]{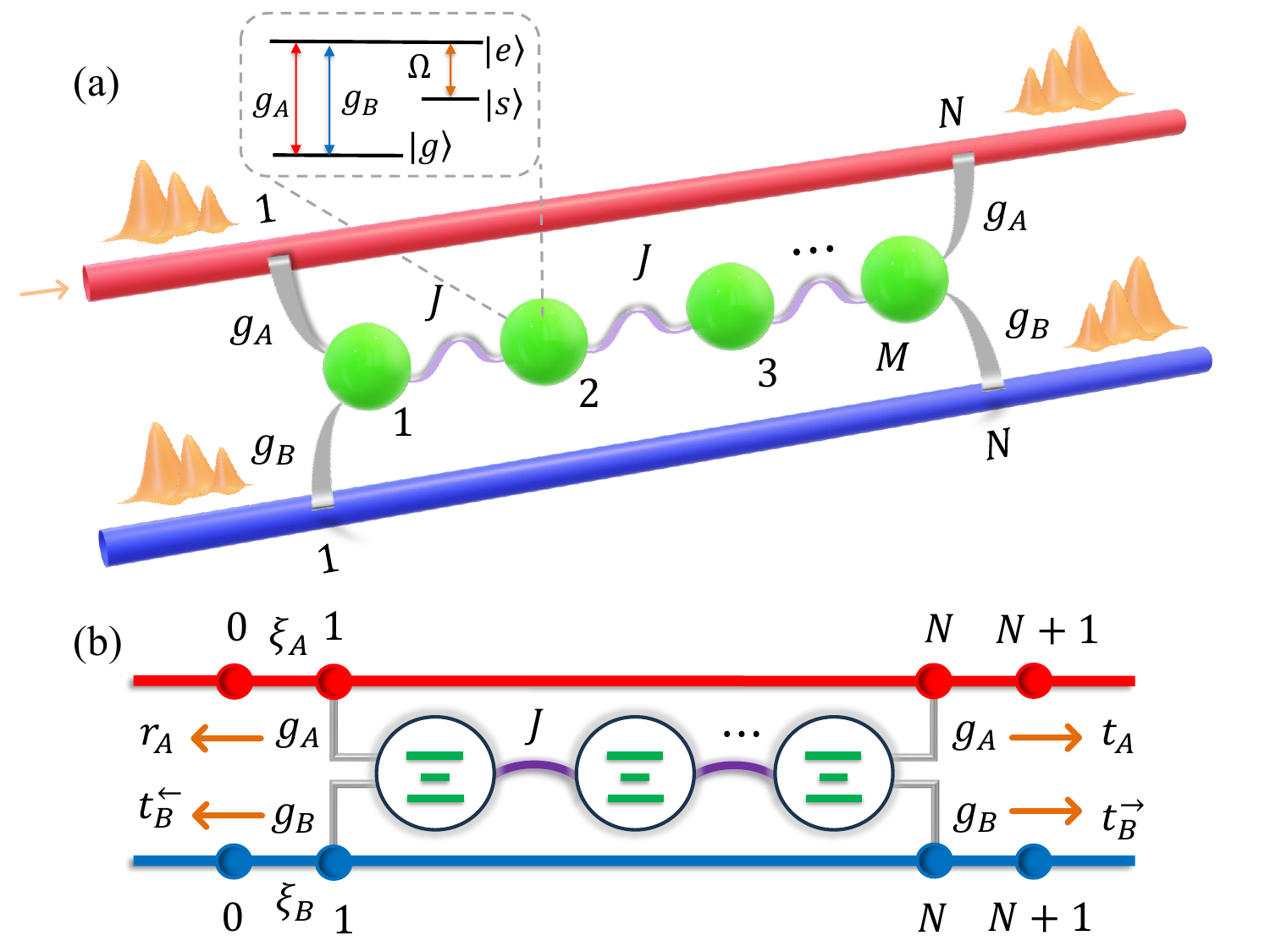}
\caption{Schematic representation of the physical system.  
(a) The upper panel illustrates the configuration of a coupled array of atoms that collectively behaves as a giant atom, interacting with two one-dimensional waveguides at spatially separated points. This configuration enables multi-path interference and non-local photon–atom coupling.  
(b) The lower panel shows a simplified schematic of the system, highlighting the effective giant-atom-like behavior arising from the spatially distributed coupling between the array and the two waveguides.}
\label{fig1}
\end{figure}

The system under consideration consists of a coupled array of atoms that collectively behaves as a giant atom, coupled to two one-dimensional waveguides. This system is modeled using a one-dimensional tight-binding lattice, as shown in Fig. \ref{fig1}. Each atom features a ground state denoted as $|g\rangle$ and an excited state represented by $|e\rangle$. 
The transition between the states $|g\rangle$ and $|e\rangle$ for each atom is dipole coupled to the waveguides $A$ and $B$, with coupling strengths denoted $g_A$ and $g_B$, respectively. The transition $|g\rangle \leftrightarrow |s\rangle$ is prohibited. An external classical control field, operating at frequency $\nu$, drives the transition between states $|s\rangle$ and $|e\rangle$ with a Rabi frequency represented by $\Omega$.
The Hamiltonian for the system is expressed as  
\begin{equation}
H = H_{AB} + H_{\text{GA}} + H_{\text{I}},
\end{equation} 
where $H_{AB}$ describes the propagation of photons through waveguides $A$ and $B$, $H_{\text{GA}}$ corresponds to the free Hamiltonian of a giant-atom-like array, and $H_{\text{I}}$ represents the interaction between the effective giant atom and the waveguides. The explicit form of the Hamiltonian is provided below, assuming $\hbar = 1$ for simplicity.


\begin{eqnarray}
H_{AB} &=&\sum_{i=-\infty}^{\infty} [ \omega_A a_i^\dagger a_i 
- \xi_A ( a_i^\dagger a_{i+1} + a_{i+1}^\dagger a_i ) ] \nonumber \\
&+& \sum_{i=-\infty}^{\infty} [ \omega_B b_i^\dagger b_i 
- \xi_B ( b_i^\dagger b_{i+1} + b_{i+1}^\dagger b_i ) ] \, ,
\label{1.a} \\  
H_{\text{GA}}&=& \sum_{l=1}^{M} [ \omega_{e}\,\sigma^{+}_{l} \sigma^{-}_{l} +\omega_{s}\,\sigma^{-}_{l, \, s} \sigma^{+}_{l, \, s} ] \nonumber \\
&+& \sum_{l=1}^{M} \Omega \, \sigma^{+}_{l, \, s} \, e^{- i\nu t} + \text{h.c.}, 
\label{1.b}  \\
H_I &=& \sigma^{+}_{1} ( g_A a_{1} + g_B b_{1} ) +  \sigma^{+}_{M} (g_A a_{N} + g_B b_{N} ) \nonumber \\
 &+&   \sum_{j=1}^{M} J \sigma^{+}_{j} \sigma^{-}_{j+1} + \text{h.c.},
\label{1.c} 
\end{eqnarray}

where $a_{i}^{\dagger}$ ($a_{i}$) and $b_{i}^{\dagger}$ ($b_{i}$) are the creation (annihilation) operators of a single photon in the $i$th waveguides sites, $A$ and $B$ with frequencies $\omega_A$ and $\omega_B$ and $\xi_{A}$ and $\xi_{B}$ is the hopping coefficient between any two nearest-neighbor sites in the waveguides, respectively. The operators $\sigma^{+}_{i} \, ( \sigma^{-}_{i})$ denote the raising (lowering) operator for the giant atom with $\sigma^{+}_{i, \, s}= | e_i \rangle \langle s_i |$ and $\sigma^{-}_{i, \, s}= | s_i \rangle \langle e_i |$. The $\omega_s$ and $\omega_e$ are the third- and excited-state frequencies, and h.c. stands for the Hermitian conjugate. The dispersion relations for the waveguides $A$ and $B$ are given by $E_A = \omega_A - 2\xi_A \cos k_A$ and $E_B = \omega_B - 2\xi_B \cos k_B$ with $k_A, \, k_B \in [0,2\pi]$, resulting in energy bands with bandwidths $4\xi_A$ and $4\xi_B$, respectively.

In this context, we examine the single-photon scattering process in the rotating frame. To facilitate analysis, we apply a unitary transformation~\cite{Ahumada2019}, given by $H' = U^{\dagger} H U - i U^{\dagger} \frac{\partial}{\partial t} U$, where $U = \prod_{j=1}^{N} e^{i \nu t |s_j\rangle \langle s_j|}$. This transformation makes the Hamiltonian time-independent, yielding $H' = H_{AB} + H'_{\text{GA}} + H'_{\text{I}}$, where
\begin{small}
\begin{eqnarray}
H'_{\text{GA}} &=& \sum_{l=1}^{M} [\omega_{e}\,\sigma^{+}_{l} \sigma^{-}_{l}   + \omega_s' \,\sigma^{-}_{l, \, s} \sigma^{+}_{l, \, s} + \Omega \, \sigma^{+}_{l, \, s} ] + \text{h.c.}, \label{1.d} \\
H'_{\text{I}} &=& \sigma^{+}_{1} ( g_A a_{1} + g_B b_{1} ) +  \sigma^{+}_{M} (g_A a_{N} + g_B b_{N} )\nonumber \\
 &+&  \sum_{j=1}^{M} J \sigma^{+}_{j} \sigma^{-}_{j+1}+ \text{h.c.}, \label{1.e}
\end{eqnarray}
\end{small}
and $\omega_s' = \omega_s + \nu$. Under this transformation, the Hamiltonian $H_{AB}$ remains unchanged.

The propagation of a single photon through the system can be assessed by inspecting the energy spectrum of the Hamiltonian. This can be obtained by expressing the single-excitation eigenstate as
\begin{eqnarray}
| \psi \rangle &=& \sum_{i=-\infty}^{\infty}  [ \alpha_{i} a_i^\dagger | 0, g \rangle + \beta_{i} b_i^\dagger | 0, g \rangle  ] \nonumber \\
&+& \sum_{j=1}^{M} [u_{e, j} | 0, e_{j} \rangle  + u_{s, j} | 0, s_{j} \rangle ] \, .
\label{1.f}
\end{eqnarray}
Here $\alpha_{i}$ and $\beta_{i}$ are the probability amplitudes to find the photon in the ith cavity of waveguides $A$ and $B$, respectively,  $u_{e, j}$ and $u_{s, j}$ are the probabilities amplitudes of the excited state and the third state, respectively, and $|0 \rangle$ is the vacuum state of the waveguides. 

\section{Transmission of a single photon in a giant-atom-like array}
\label{sec:2}
Our proposed system consists of two waveguides, each forming a quasi-one-dimensional array of identical optical cavities with nearest-neighbor coupling. 
The coupled stationary equations for the amplitudes are derived from the eigenvalue equation $H |\psi \rangle = E |\psi \rangle$
\begin{small}
\begin{subequations}
\begin{eqnarray}
(E - \omega_e) u_{e,j} &=& J (u_{e,j+1} + u_{e,j-1}) +\Omega u_{s,j} \nonumber \\
&&  \hspace{-2cm} + g_{A} (\delta_{j,1} \alpha_1 + \delta_{j,M} \alpha_{N}) + g_{B} (\delta_{j,1} \beta_1 + \delta_{j,M}\beta_{N}) \, ,  \label{3.a} \\
(E - \omega'_s) u_{s,j} &=& \Omega^{*} u_{e,j} \, ,  \label{3.b} \\
(E - \omega_A) \alpha_i &=& -\xi_A (\alpha_{i+1} + \alpha_{i-1})\nonumber \\
&+& g_{A} \sum_{j} \,  (\delta_{i,1} \delta_{j,1} +\delta_{i,N} \delta_{j,M}) \, u_{e,j} \, , \label{3.c} \\
(E - \omega_B) \beta_i &=& -\xi_B (\beta_{i+1} + \beta_{i-1}) \nonumber\\
&+& g_{B} \sum_{j} \,  (\delta_{i,1} \delta_{j,1} +\delta_{i,N} \delta_{j,M}) \, u_{e,j} \, , \label{3.d} 
\end{eqnarray}
\end{subequations}
\end{small}

From (\ref{3.a}) and (\ref{3.b}) we obtain the coupled equations,
\begin{equation}
(E-\bar{\omega}_{e})u_{e,j} - J (u_{e,j+1} + u_{e,j-1}) = S_{j}  ,\label{eq.4a}
\end{equation}
with $\bar{\omega}_{e}= \omega_{e}+\Omega^2 / (E-\omega'_s)$ and $S_{j}= g_{A} (\delta_{j,1} \alpha_1 + \delta_{j,M} \alpha_{N}) + g_{B} (\delta_{j,1} \beta_1 + \delta_{j,M}\beta_{N})$. This results in a nonhomogeneous equation for the chain of atoms coupled through $J$. We can thus solve Eq.~(\ref{eq.4a}) using Green's functions, obtaining the solution $u_{e,j} = \sum_{j} \, S_{j'}\, \langle j | G(E) | j' \rangle$. Once $u_{e,j}$ has been calculated, this solution is substituted in expressions (\ref{3.c}) and (\ref{3.d}), obtaining
\begin{footnotesize}
\begin{subequations}
\begin{eqnarray}
(E - \omega_A ) \alpha_i &=& -\xi_A (\alpha_{i+1} + \alpha_{i-1}) \nonumber \\
&&\hspace{-7em}+  \Gamma(E) (\delta_{i,1} +\delta_{i,N}) \left[ g_A^2 (\alpha_{1} +\alpha_{N}) + g_Ag_B (\beta_{1} + \beta_{N}) \right] , \label{eq.4b} \\
(E - \omega_B ) \beta_i &=& -\xi_B (\beta_{i+1} + \beta_{i-1}) \nonumber \\
&&\hspace{-7em}+  \Gamma(E) (\delta_{i,1} +\delta_{i,N}) \left[ g_Ag_B (\alpha_1 + \alpha_{N}) + g_B^2 (\beta_{1} +\beta_{N})  \right] , \label{eq.4c} 
\end{eqnarray}
\end{subequations}
\end{footnotesize}

with $\Gamma(E) = \langle 1 | G(E) | 1 \rangle +\langle 1 | G(E) | M \rangle = \langle M | G(E) | 1 \rangle +\langle M | G(E) | M \rangle $ (see Appendix \ref{App.A}).

For a plane wave incident from $-\infty$ in the waveguide, under the standard scattering boundary conditions, the photon amplitudes in the two channels can be written as 
\vspace{-4mm}
\begin{equation}
    \begin{aligned}
        \alpha_j &=
        \begin{cases}
            e^{ik_A j} + r_A e^{-ik_A j}, & j < 1 \\
            A e^{ik_A j} + B e^{-ik_A j}, & 1 \leq j \leq N \\
            t_A e^{ik_A j}, & j > N \, ,
        \end{cases} \\
        \beta_j &=
        \begin{cases}
            t_B^{\leftarrow} e^{-ik_B j}, & j < 1 \\
            C e^{ik_B j} + D e^{-ik_B j}, & 1 \leq j \leq N \\
            t_B^{\rightarrow} e^{ik_B j}, & j > N \label{5} \, ,
        \end{cases}
    \end{aligned}
\end{equation}
where $r_A$ and $t_A$ are the reflection and transmission amplitudes in channel $A$, and $t^B_{\leftarrow}$ and $t^B_{\rightarrow}$ are the backward and forward transfer amplitudes in channel $B$, respectively.

 To ensure maximum overlap between the energy bands of the two waveguides, we set $\omega_A = \omega_B = \omega_0$ and $\xi_A = \xi_B = \xi$. Throughout this paper, the nearest-neighbor coupling $\xi$ will be used as the unit of energy. Additionally, we assume equal giant-like atom array -to-waveguide mode couplings, given by $g_{A} = g_{B} = g$. 

To solve Eqs.~(\ref{eq.4b}) and (\ref{eq.4c}), we perform the following transformation: Instead of considering the photon amplitudes $\alpha_j$ and $\beta_j$ with $j \in \{1, N\}$ in the physical waveguides $A$ and $B$, we introduce their symmetric and anti-symmetric combinations for each waveguide. Specifically, we define: $\psi^{\pm}_{j} = \alpha_j \pm \beta_j$. In this representation, the equations are simplified, leading to decoupled symmetric ($\mathcal{S}$) and antisymmetric ($\mathcal{A}$) modes for the waveguides $A$ and $B$.

\begin{figure}[h]
\includegraphics[width=\columnwidth]{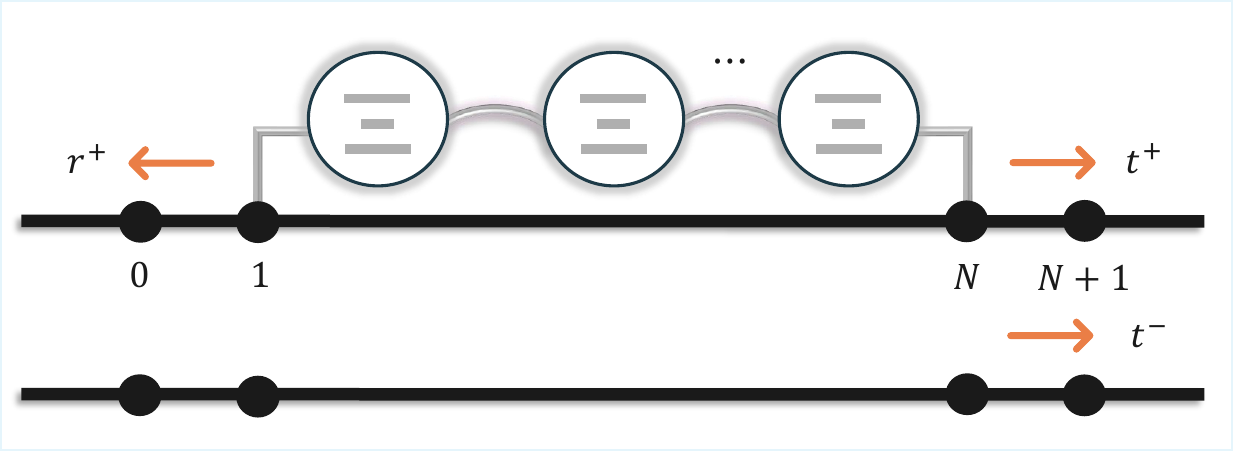}
\caption{Schematic representation of the virtual symmetric and antisymmetric channels for a giant-atom-like array. These channels are labeled with $+$ and $-$, respectively. The amplitudes of the transmission $t^{+}$ and $t^{-}$ and reflection $r^{+}$ are represented by orange wavy arrows.}
\label{fig3}
\end{figure}

The equations for symmetric and antisymmetric representation
\begin{small}
\begin{equation}
    \begin{aligned}
        (E - \varepsilon^{+}) \psi_j^+ &= -\xi (\psi_{j+1}^+ + \psi_{j-1}^+) +  \bar{\varepsilon}( \psi_{N}^+ \delta_{j, 1} +\psi_{1}^+ \delta_{j, N}) \, , \\
        (E - \varepsilon^{-}) \psi_j^- &= -\xi (\psi_{j+1}^- + \psi_{j-1}^-) \label{6} \, ,
    \end{aligned}
\end{equation} 
\end{small}
where the effective site energies are $\varepsilon^{+}=\omega_{0} +\bar{\varepsilon}$, $\varepsilon^{-}=\omega_0$ and $\bar{\varepsilon}= g^2 \Gamma(E)$.

In the $\mathcal{S}\text{-}\mathcal{A}$ representation, the scattering boundary conditions are written as
\begin{small}
\begin{equation}
    \begin{aligned}
        \psi_j^{\pm} &=
        \begin{cases}
            e^{ik_{\pm} j}  + r^{\pm} e^{-ik_{\pm} j} , & j < 1 \\
            A^{\pm} e^{ik_{\pm} j} + B^{\pm} e^{-ik_{\pm} j}, & 1 \leq j \leq N \\
            t^{\pm} e^{ik_{\pm} j}, & j > N \, .
        \end{cases} 
        \label{7} 
    \end{aligned}
\end{equation}
\end{small}

The transmission amplitudes ($t^{\pm}$) and the reflection ($r^{\pm}$) correspond to the virtual channels $\mathcal{S}\text{-}\mathcal{A}$. From Eqs. (\ref{6}), evaluated at the boundaries of the scattering region $j=\{0, 1, N, N+1\}$, together with Eqs. (\ref{7}) (see Appendix \ref{App.B}), we can derive closed-form expressions for $t^{\pm}$ and $r^{\pm}$. Since the $\mathcal{A}$ channel behaves as a free channel with energy $\omega_0$, the incident wave is fully transmitted without reflection, resulting in a unit transmission amplitude: $r^{-} = 0$ and $t^{-} = 1$. The values of $r^{+}$ and $t^{+}$ can then be determined accordingly. Specifically, the reflection and transmission amplitudes for the $\mathcal{S}$ channel, as well as the wave number $k_{+}$, are given by:
\vspace{-1.5em}
\begin{footnotesize}
\begin{subequations}
\begin{align}
r^+ &= \frac{4 \xi \sin k}{\Lambda(k_+)} \left[ \Delta(k_+) - \chi(k_+) \right], \label{9.a} \\
t^+ &= \frac{4 \xi^3 \sin k \, e^{-i k (N+1)}}{\Lambda(k_+)} \left[ \bar{\varepsilon} \sin\big( k_+ (N+1) \big) - \xi \sin k_+ \right], \label{9.b} 
\end{align}
\end{subequations}    
\end{footnotesize}
with $k_+ = \arccos ( -(E - \varepsilon^{+}) / 2\xi)$, if $|(E - \varepsilon^{+})/2\xi| \leq 1$ and
\begin{footnotesize}
\begin{subequations}
\begin{align}
\chi(k_+) &= \xi \sin\big( k_+ (N-2) \big) -2 \xi \cos k_{+} \, \sin\big( k_+ (N-1) \big), \label{9.c} \\
\lambda(k_{+})&= 2i \xi^2 (E - \omega_0 + \xi e^{i k_+}) \chi(k_+),  \label{9.d} \\
\Delta(k_+) &= \left[ 4 \xi^2 \cos^2 k_{+} - (\bar{\varepsilon} + \xi)^2 \right] \sin\big( k_+ (N-1) \big) \nonumber\\
& \hspace{-3.8em} +\sin\big( 2 k_+ (N-1) \big) \left[  2 \xi (\bar{\varepsilon} - \xi) \cos k_{+} \, + \xi^2 \cos\big( k_+ (N-1) \big) \right], \label{9.e} \\
 \Lambda(k_+) &= -\Delta(k_+) - 2i \xi^4 \sin\big( k_+ (N-1) \big) + \lambda(k_{+}). \label{9.f} 
\end{align}
\end{subequations}    
\end{footnotesize}

Once the transmission and reflection amplitudes in the virtual waveguides $\mathcal{S}$ and $\mathcal{A}$ are known, the corresponding amplitudes for the physical waveguides $A$ and $B$ can be obtained by applying the inverse transformation, which corresponds to the inverse of the transformation that decouples the amplitudes into symmetric and antisymmetric components. From this, the transmission and reflection amplitudes for the $A$ and $B$ waveguides can be derived:
\vspace{-1mm}
\begin{small}
\begin{equation}
r_A = t_B^{\leftarrow} = \frac{1}{2} r^+ \, , \quad
t_A = \frac{1}{2} (t^+ + 1) \, , \quad 
t_B^{\rightarrow} = \frac{1}{2} (t^+ - 1) \, .
\label{eq.10}
\end{equation}
\end{small}

The reflection, transmission and transfer probabilities are calculated as $R_A = |r_A|^2$, $T_A = |t_A|^2$, $T_B^{\leftarrow} = |t_B^{\leftarrow}|^2$, and $T_B^{\rightarrow} = |t_B^{\rightarrow}|^2$. These scattering amplitudes satisfy the standard flow conservation condition: $R_A + T_A + T_{B}^{\leftarrow} + T_{B}^{\rightarrow} = 1$.

The model in the $\mathcal{S}$–$\mathcal{A}$ representation is schematically illustrated in Fig.~\ref{fig3}. The $\mathcal{S}$ waveguide is analogous to a conventional waveguide with embedded two-level atoms. Importantly, the $\mathcal{S}$ and $\mathcal{A}$ waveguides are decoupled. In the symmetric $\mathcal{S}$ waveguide, the effective site energy $\varepsilon^{+}$ is renormalized with respect to $\omega_0$ within the scattering region. Additionally, a giant-atom-like array is connected to sites $1$ and $N$, inducing scattering processes within this waveguide. In contrast, the site energy in the $\mathcal{A}$ waveguide remains uniform, allowing for free wave propagation.

\section{Results}
\label{Results}


In what follows, we use $\xi$ as the unit of energy. The system configuration is illustrated in Fig.~\ref{fig1}, where the device functions as an optimal quantum router, achieving maximum transfer probability. This optimal behavior is observed when the coupling between waveguides~$A$ and~$B$ and the effective giant atom is strong, specifically when $g = g_A = g_B = 1.5$. In contrast, the coupling within the array of atoms that collectively behave as a giant atom is weak, with $J = 0.01$ (all quantities expressed in units of $\xi$).

In this approach, we examine the simplest scenario with zero control field ($\Omega=0$), which results in the decoupling of the third states $| s_j \rangle $ from the rest of the system. In this setup, an incident photon is scattered by a system of waveguides that are coupled through $N$ two-level atoms. In particular, the number of atoms $M$ equals the number of coupling sites $N$ between the waveguides and the array that behaves as a giant atom. This condition is crucial for achieving the desired interference and is schematically illustrated in Fig.~\ref{fig1}.

\begin{figure}[h]
\includegraphics[width=\columnwidth]{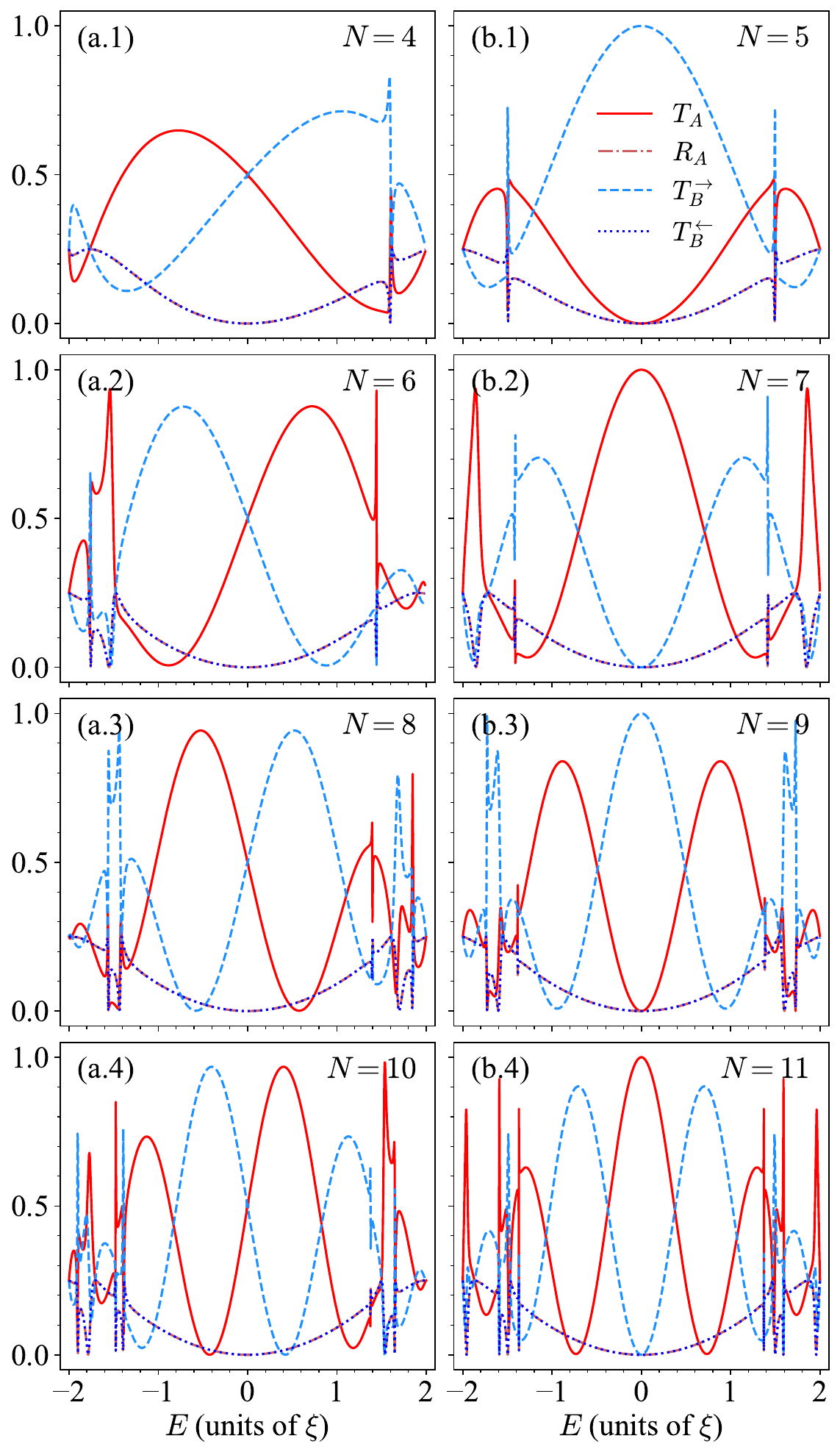}
\caption{Transmission probability $T_A$ (solid red
line), reflection $R_A$ (red dashed line), and transfer $T^{\rightarrow}_B$ (blue dotted-and-dashed line) and $T^{\leftarrow}_B$ (blue dotted line) spectra as a function of the incident energy $E$. The spectra are calculated for the parameters $\omega_0 = \omega_s = \omega_e = 0$, $\Omega = 0$, $J = 0.01$, and $g = 1.5$ (all units of $\xi$), for different numbers of atoms as indicated in the panels~(a.\,1)--(a.\,4) and~(b.\,1)--(b.\,4).}
\label{fig5}
\end{figure}

Fig.~\ref{fig5} displays the transmission, reflection and photon transfer probabilities as functions of the incident energy $E$  for the case where $\omega_0 = \omega_s = \omega_e = 0$, $\Omega = 0$, $J = 0.01$, $g = 1.5$ (all units of $\xi$), and the varying numbers of atoms $N$. 
In the subpanels labeled (a.1) to (a.4), which correspond to even values of $N$, the router shows optimal performance, achieving photon transfer probabilities close to $100\%$  across a wide energy range centered around $E = 0$. In contrast, the subpanels (b.1) to (b.4), related to odd values of $N$, reveal that the maximum photon transfer does not occur at the center of the band.
The most effective configurations are found in Fig.~\ref{fig5}, particularly in (a.3) with $N = 8$ and (a.4) with $N = 10$, as well as in (b.1) with $N = 5$ and (b.3) with $N = 9$, where high photon transfer is also observed near the center of the band.
\begin{figure}[h]
\includegraphics[width=\columnwidth]{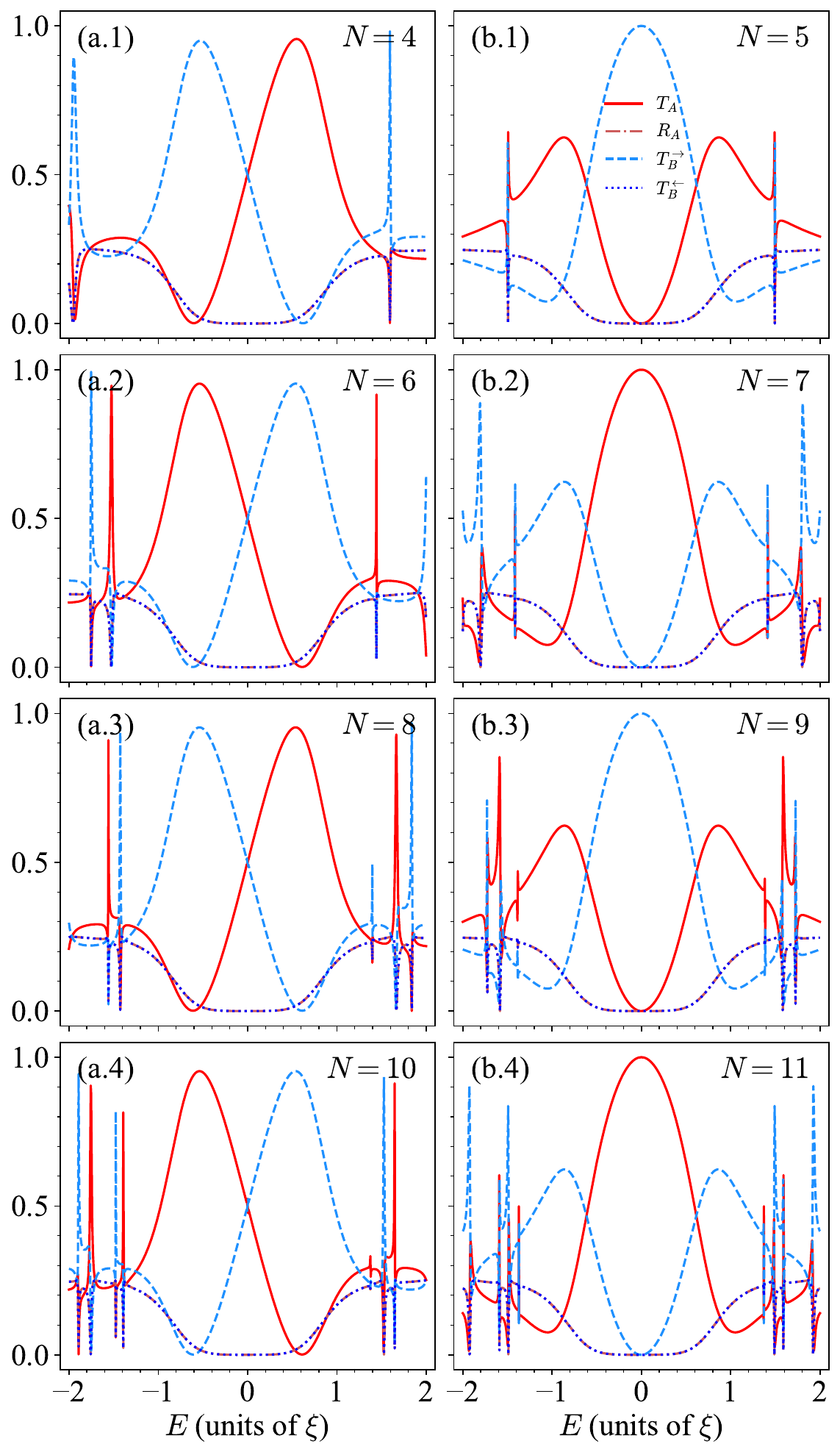}
\caption{Transmission probability $T_A$ (solid red
line), reflection $R_A$ (red dashed line), and transfer $T^{\rightarrow}_B$ (blue dotted-and-dashed line) and $T^{\leftarrow}_B$ (blue dotted line) spectra as a function of the incident energy $E$. The spectra are calculated for the parameters $\omega_0 = \omega_s = \omega_e = 0$, $\Omega = 0$, $J = 0.01$, $g = 1.5$ (all units of $\xi$), and $k = \pi /2$.}
\label{fig5.1}
\end{figure}
Furthermore, we can observe a distinct periodic pattern, with the transfer probabilities approaching $100\%$ near the center of the energy band. To examine the energy-dependent periodicity of the router, we fix the wave vector at the center of the band ($k = \pi/2$) in the analytical expressions for $r_+$ and $t_+$ [Eqs.~(\ref{9.a}) and (\ref{9.b})].  As illustrated in Fig.~\ref{fig5.1}, the resulting periodicity is $\tau = 4N$  for any number of atoms. Furthermore, when $N$ is even, this periodicity is centered within the energy band [panels (b.1)–(b.4)]. However, for odd $N$, the periodicity is no longer centered within the band [panels (a.1)–(a.4)].

To understand the periodic pattern observed in the transmission spectra presented in the results of Fig.~\ref{fig5.1}, we analyze the spectral periodicity of the transmission amplitude $t^+$ [Eq.~(\ref{9.b})] for a fixed $N$. We focus on the regime where $k = \pi/2$, corresponding to the center of the energy band. The system is described in natural units ($\hbar = 1$, $\xi = 1$), with energy measured in the interval $E \in [-2 , 2]$, and the wavevector in the scattering region defined as $k_+(E) = \arccos(-E/2)$.

Under the condition that the self-energy correction $\bar{\varepsilon}(E) = g^2 \Gamma(E) \approx 0$ near $E = 0$, the transmission amplitude simplifies to $t^+ \approx \frac{-4  \sin k_+ }{\Lambda(k_+(E))},$
where the denominator $\Lambda(k_+(E))$ [Eq. (\ref{9.f})], involves trigonometric functions of the form $\Lambda(k_+(E)) \sim f (k_+(N-j))$ with $j=1,2$. The spectral structure of $t^+$ is governed by the oscillatory behavior of the phase
$\phi(E) = k_+(E) \cdot N$, which corresponds to the accumulated phase of the photonic wavefunction between the two distant coupling sites $1$ and $N$. 

In quantum scattering theory, the wave function of the incident photon acquires a phase proportional to the distance between scattering centers. In this case, the photonic state within the waveguide can be written (up to normalization) as:
\begin{equation}
\psi_j(E) \sim e^{i k_+(E) j},
\label{11}
\end{equation}
and when interacting with two distant sites $1$ and $N$, the phase difference accumulated between them is
\begin{equation}
\phi(E) = k_+(E) (N - 1),
\label{12}
\end{equation}
which governs interference. This is called \textit{the effective phase}, and plays a central role in determining constructive or destructive interference, and thus resonances or suppressions in the transmission amplitude $t^{+}$.

The key periodicity arises when this phase increases by $2\pi$, i.e., $\phi(E + \Delta E) = \phi(E) + 2\pi$
\begin{equation}
k_+(E + \Delta E) = k_+(E) + \frac{2\pi}{N}.
\label{13}
\end{equation}

The dispersion relation $E = -2 \cos(k_+(E))$, and thus, a shift in phase corresponds to an energy shift
\begin{equation}
\Delta E = -2 \left[ \cos\left(k_+(E) + \frac{2\pi}{N} \right) - \cos(k_+(E)) \right].
\label{14}
\end{equation}
For $N \gg 1$, a Taylor expansion gives
\begin{equation}
\Delta E \approx \frac{4\pi}{N} \sin(k_+(E)) + \frac{4\pi^2}{N^2} \cos(k_+(E)),
\label{15}
\end{equation}
which shows that the spectral periodicity depends on $E$ through $k_+(E)$, but around $E \approx 0$, where $k_+(E) \approx \pi/2$, the dominant term yields
\begin{equation}
\Delta E \approx \frac{4\pi}{N}.
\label{16.a}
\end{equation}

However, in natural units used in numerical simulations, where $\pi \equiv 1$, this simplifies numerically to
\begin{equation}
\Delta E = \frac{4}{N},
\label{16.b}
\end{equation}
and the full transmission pattern as a function of $E$ becomes exactly periodic every $4N$ units of energy, i.e., $\tau=4N$.

Therefore, the effective phase $\phi(E) = k_+(E) N$, resulting from the accumulated phase between sites $1$ and $N$, explains the periodic behavior of $t^+$. Using the relations $t_A = \frac{1}{2}(t^+ + 1)$ and $t_B^{\rightarrow} = \frac{1}{2}(t^+ - 1)$ [Eqs.~(\ref{eq.10})], this behavior is reflected in the transmission probabilities $T_A = |t_A|^2$ and $T_B^{\rightarrow} = |t_B^{\rightarrow}|^2$ and the spectral repetition observed in Fig.~\ref{fig5.1}. The result agrees well with numerical simulations in natural units.

\begin{figure}[h]
\includegraphics[width=\columnwidth]{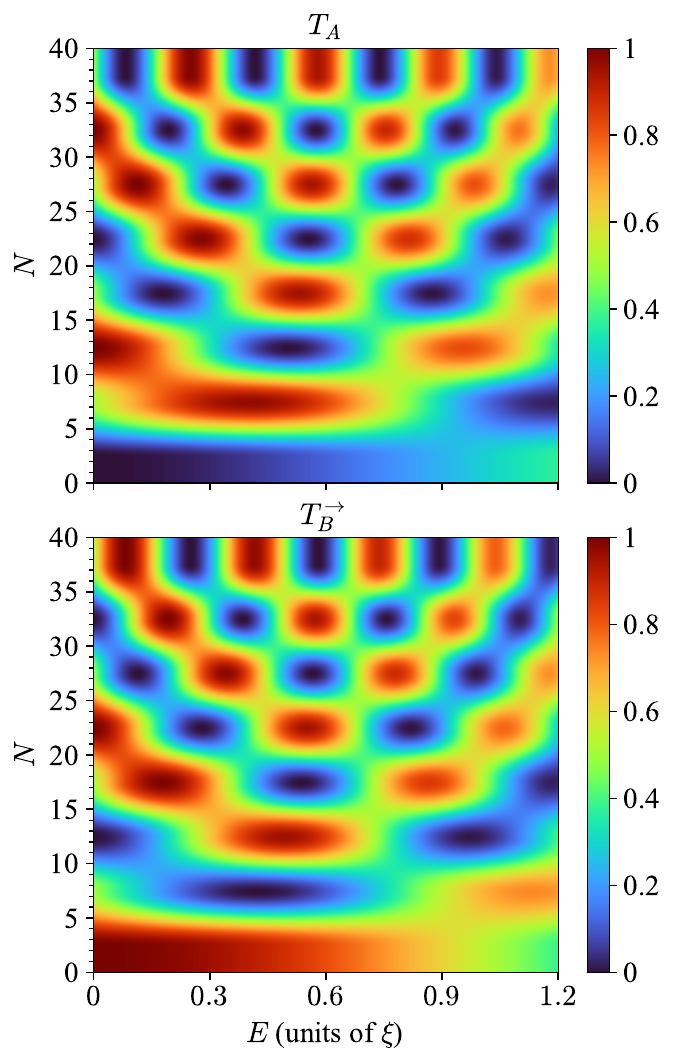}
\caption{Density Probability transmission $T_A$ and transfer $T_B^{\rightarrow}$ spectra as a function of the incident energy $E$ and the number of atoms $N$. The spectra are calculated for the parameters $\omega_0 = \omega_s = \omega_e = 0$, $\Omega = 0$, $J = 0.01$, and $g = 1.5$ (all units of $\xi$).}
\label{fig6}
\end{figure}

We now explore a system configuration specifically designed to maximize photon routing performance, focusing on the central region $E \sim [0, 1.2]$. Fig.~\ref{fig6} shows the transmission and transfer spectra as functions of the photon energy $(E > 0)$, and the number of atoms $N$, corresponding to the transmission probability $T_A$ and the transfer probability $T_B^{\rightarrow}$. The parameters used are $\omega_0 = \omega_s = \omega_e = 0$, $\Omega = 0$, $J = 0.01$, and $g = 1.5$ (all units of $\xi$).

In Fig.~\ref{fig6}, we observe contrasting behaviors between $T_A$ and $T_B^{\rightarrow}$. Although $T_A$ exhibits regions of higher intensity, the corresponding values of $T_B^{\rightarrow}$ are significantly suppressed in these same regions. Focusing on the transfer probability $T_B^{\rightarrow}$, a repeating pattern as a function of $N$ becomes apparent when the energy is centered within the band, with $T_B^{\rightarrow} = 1$ occurring periodically. In particular, the case of $N=20$  atoms corresponds to five full periods, i.e., $5\tau$.

\begin{figure*}[ht]
\includegraphics[width=\textwidth]{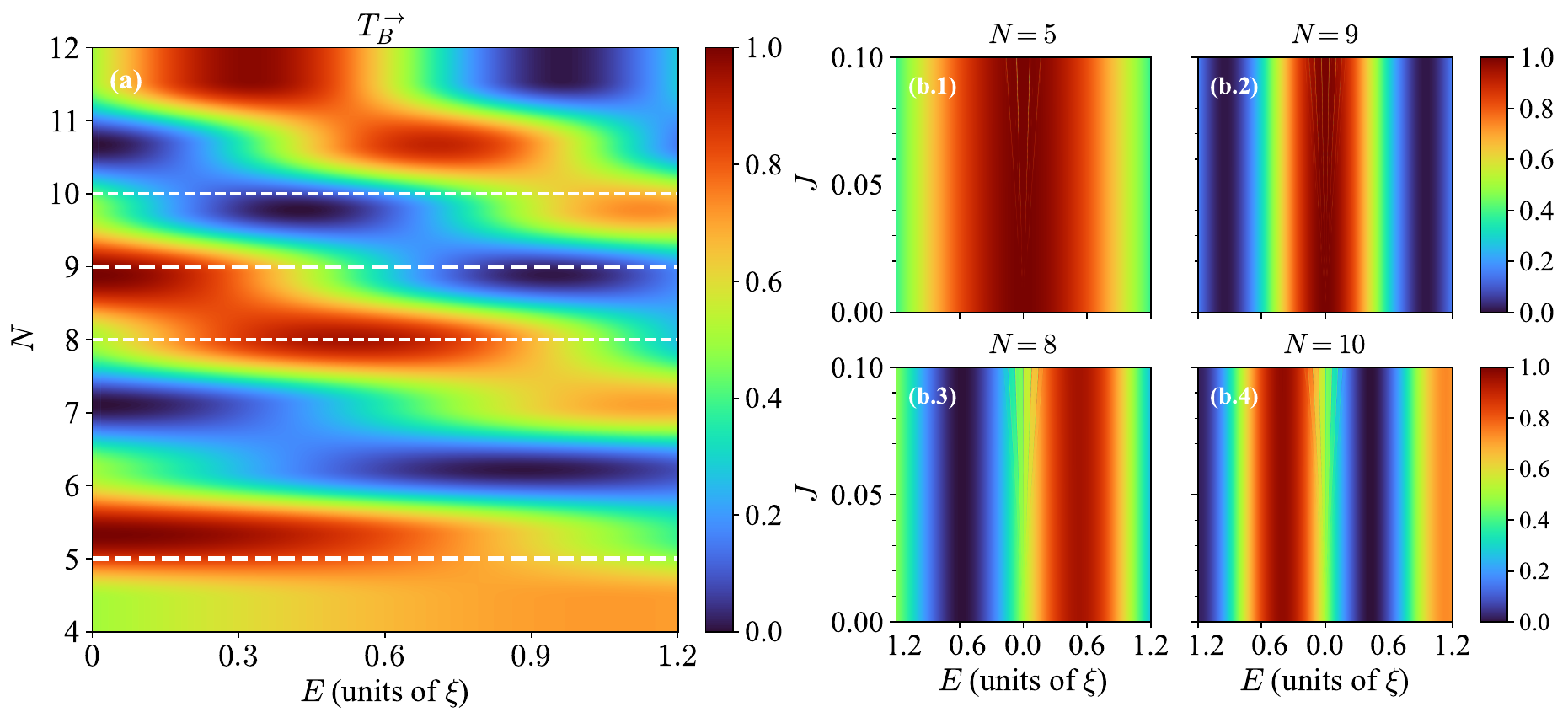}
\caption{Panel~(a) shows the density plot of the transfer probability $T_B^{\rightarrow}$ as a function of the incident energy $E$ and the number of atoms $N$. The white dashed line indicates the optimal system configuration
of $N=5$, $N=8$, $N=9$, and $N=10$. The spectra are calculated for the parameters $\omega_0 = \omega_s = \omega_e = 0$, $\Omega = 0$, $J = 0.01$, and  $g = 1.5$. The right panels display the transfer probabilities $T_B^{\rightarrow}$ as a function of the incident energy $E$ and the coupling strength $J$ of the atoms. The spectra are calculated for the parameters $\omega_0 = \omega_s = \omega_e = 0$, $\Omega = 0$, and $g = 1.5$ (all units of $\xi$), for (b.\,1) $N=5$, (b.\,2) $N=9$, (b.\,3) $N=8$, and (b.\,4) $N=10$.}
\label{fig7}
\end{figure*}

To understand the conditions for perfect routing ($T_B^{\rightarrow} = 1$),  we examine panel~(a) of Fig.~\ref{fig7}. The dashed white lines indicate the first values of $N$ at which perfect transmission is observed, with prominent peaks at $N = 5$, $8$, $9$, and $10$. Panels~(b.1) and~(b.2) in Fig.~\ref{fig7} further illustrate the behavior of $T_B^{\rightarrow}$ as a function of the incident energy $E$ and the coupling strength $J$ of the atoms. When the energy lies near the center of the band, the transmission range is approximately $E \sim [-1, 1]$ for $N = 5$ (b.1), and narrows to $E \sim [-0.5, 0.5]$ for $N = 9$ (b.2), expressed in units of $\xi$. This behavior reflects the sensitivity of the resonance to both $N$ and the energy detuning. On the other hand, panels~(b.3) and~(b.4) in Fig.~\ref{fig7} show that the energy range is shifted to $E \sim [0, 1]$ for (b.3) and $E \sim [-1, 0]$ for (b.4).

\section{Summary}
\label{sec:summary}

In this article, we analyze the transmission spectrum of a system consisting of a giant-atom-like array coupled to two waveguides. In the regime of strong atom–waveguide coupling and weak inter-atomic coupling,  we find that our setup operates as an optimal quantum router. Additionally, we observed that when the system functioned as an efficient router, it exhibited a characteristic periodicity of $\tau = 4N$.  This periodicity was linked to the accumulated phase of the photonic wave function between the two distant coupling sites, $1$ and $N$. The phase difference accumulated between these sites,
governed the interference phenomena and was referred to as the \textit{effective phase}. This phase played a central role in determining whether interference was constructive or destructive and, consequently, whether transmission resonances or suppressions occurred. 
Constructive interference appeared for even values of $N$, while destructive interference was observed for odd $N$.


The interference pattern was observed in both the transmission and transfer probability densities, which displayed a distinct periodic behavior.  When one channel reached its maximum, the other simultaneously hit its minimum. This alternating pattern occurred every five periods ($5\tau$), corresponding to a total of $N = 20$  atoms. Therefore, we focus on the most significant values within this periodicity, particularly in the transfer probability for channel $B$. 
This periodic and alternating behavior provides a mechanism for controlling the routing direction by tuning either the number of atoms or the photon energy, enabling energy-selective and configuration-dependent routing within a quantum photonic network~\cite{Reiserer2015,Wei2022,Abane2025,Li2022}.

The proposed configuration demonstrates improved performance, featuring higher routing efficiency and a broader operational bandwidth compared to previous photon routing schemes based on giant atoms ~\cite{Zhang_Yu2023,Zheng2024,Wang2025}. Notably, we identified configurations that achieve nearly $100\% $ photon transfer across a wide energy range around the center of the band. This result highlights the stability and effectiveness of the routing mechanism. These enhancements make the system less susceptible to parameter fluctuations and more stable against external disturbances, which is a significant advantage for practical experiments ~\cite{Wang2021,Li2024,Sollner2015,Zhou2015,Wang2022Chiral}. In general, our findings emphasize the reliability and versatility of our setup, reinforcing its potential as a foundational component for scalable quantum information technologies.

\acknowledgments
A.R.L. acknowledges financial support from ANID-Subdirecci\'{o}n de Capital Humano Doctorado Nacional Grant 2023-21230847. M.~M. acknowledges financial support from ANID Postdoctoral FONDECYT Grant No. 3240726. P.A.O. acknowledges support from DGIIE USM PI-LIR-24-10, FONDECYT Grants No. 122070, 1230933. We are grateful to Manuel Pino García (Universidad de Salamanca) for valuable suggestions that enhanced this work.

\appendix
\section{A Detailed Theoretical Analysis of solutions for $\Gamma(E)$}
\label{App.A}
In this appendix, we obtain the solution of the non-homogeneous equation for the finite chain of coupled giant atoms and from this solution we obtain the term $\Gamma(E)$.

The resulting equation~(\ref{eq.4a}) admits the solution $u_{e,j} = \sum_{j} \, S_{j'}\, \langle j | G(E) | j' \rangle$, where $G(E)$ denotes the Green's function associated with the discrete tight-binding operator acting on $u_{e,j}$. This function is defined as
\begin{equation}
G(E) = \frac{1}{(E - \bar{\omega}_{e}) I - J H_{\text{TB}} + i \eta},
\label{A.1}
\end{equation}
where $H_{\text{TB}}$ is a tridiagonal matrix representing a tight-binding chain with $M$ sites, nearest-neighbor hopping amplitude $J$, and onsite energy $\bar{\omega}_e$. The matrix elements $\langle j | G(E) | j' \rangle$ correspond to the resolvent of this operator.

To evaluate $\langle j | G(E) | j' \rangle$, we use its \textit{Lehmann representation}, 
\begin{equation}
G(E) = \sum_{n=1}^{M} \frac{|\psi_n\rangle \langle \psi_n|}{E - \bar{\omega}_e - J E_n + i \eta},
\label{A.2}
\end{equation}
where $H_{\text{TB}} |\psi_n\rangle = E_n |\psi_n\rangle$. Projecting onto the site basis yields
\begin{equation}
\langle j | G(E) | j' \rangle = \sum_{n=1}^{M} \frac{\langle j | \psi_n \rangle \langle \psi_n |j' \rangle}{E - \bar{\omega}_e - J E_n + i \eta},
\label{A.3}
\end{equation}
where $\langle j | \psi_n \rangle = \sqrt{\frac{2}{M+1}} \sin\left( \frac{n j \pi}{M+1} \right)$ are the normalized eigenfunctions and $E_n = 2 \cos\left( \frac{n \pi}{M+1} \right)$ are the corresponding eigenvalues.


Using this spectral decomposition, we find that $\Gamma(E)$ can be expressed as
\begin{equation}
\begin{aligned}
\Gamma(E) &= \langle 1 | G(E) | 1 \rangle + \langle 1 | G(E) | M \rangle  \\
&= \langle M | G(E) | 1 \rangle + \langle M | G(E) | M \rangle.
\end{aligned}
\label{A.4}
\end{equation}

The equality between these two expressions follows from the spatial symmetry of the finite tight-binding chain, which implies $\langle 1 | G(E) | 1 \rangle = \langle M | G(E) | M \rangle$ and $\langle 1 | G(E) | M \rangle = \langle M | G(E) | 1 \rangle.$

\section{A Detailed Theoretical Analysis of Transmission and Reflection Amplitudes}
\label{App.B}
This Appendix demonstrates how, using the scattering boundary equations, we calculate the reflection and transmission amplitudes for different configuration of the giant atoms in the symmetric ($\mathcal{S}$) and antisymmetric ($\mathcal{A}$) representations.

The transformation that decoupled symmetric ($\mathcal{S}$) and antisymmetric ($\mathcal{A}$) modes for the waveguides $A$ and $B$,
\begin{equation}
\begin{pmatrix}
\psi^{+}_{j} \\
\psi^{-}_{j}
\end{pmatrix}
= S
\begin{pmatrix}
\alpha_{j} \\
\beta_{j}
\end{pmatrix}
\quad \text{with } \quad S = \begin{pmatrix}
1 & 1 \\
1 & -1 
\end{pmatrix}\, ,
\label{B0.1}
\end{equation}
the corresponding amplitudes for the physical waveguides $A$ and $B$ can be obtained by applying the inverse transformation,
\begin{equation}
\begin{pmatrix}
\alpha_{j} \\
\beta_{j}
\end{pmatrix}
= S^{-1}
\begin{pmatrix}
\psi^{+}_{j} \\
\psi^{-}_{j}
\end{pmatrix}
\quad \text{with } \quad S^{-1} = \frac{1}{2}\begin{pmatrix}
1 & 1 \\
1 & -1 
\end{pmatrix}\, .
\label{B0.2}
\end{equation}
The matrix $S$ is symmetric and invertible, and satisfies the relation $S S^{-1} = I$. We then apply the inverse transformation (Eq.~\ref{B0.2}) to express the reflection and transmission amplitudes in the $\mathcal{S}\text{-}\mathcal{A}$ basis back in terms of the physical amplitudes associated with waveguides $A$ and $B$. This yields the reflection, transmission, and transfer amplitudes in the original waveguide representation.

\subsection{Coupled giant atoms system}
In this part, we use the Eqs.~(\ref{6}) with the continuous condition at $j=\{0, 1, N, N+1\}$, which is
\begin{subequations}
\begin{align}
    (E -\omega_{0}) \psi_{0}^{\pm} &= \xi (\psi_{1}^{\pm} +\psi_{-1}^{\pm}) \, , \label{B1.a} \\
    (E -\varepsilon^{\pm }) \psi_{0}^{\pm} &= \xi (\psi_{2}^{\pm} +\psi_{0}^{\pm}) +\bar{\varepsilon} \psi_{N}^{\pm} \, , \label{B1.b} \\
    (E -\varepsilon^{\pm }) \psi_{N}^{\pm} &= \xi (\psi_{N+1}^{\pm} +\psi_{N-1}^{\pm}) +\bar{\varepsilon} \psi_{1}^{\pm} \, , \label{B1.c} \\
    (E -\omega_{0}) \psi_{N+1}^{\pm} &= \xi (\psi_{N+2}^{\pm} +\psi_{N}^{\pm}) \, . \label{B1.d}
\end{align}
\end{subequations}
In the scattering region ($j < 1$ or $j > N$), the parameters take the values $\varepsilon^{+} = \varepsilon^{-} = \omega_0$ and $k_+ = k_- = k$. Under these conditions, we combine Eqs.~(\ref{B1.a})--(\ref{B1.d}) with Eqs.~(\ref{7}). Solving this system yields Eqs.~(\ref{9.a}) and~(\ref{9.b}). Finally, applying the inverse transformation Eq.~(\ref{B0.2}), 
\begin{equation}
\begin{pmatrix}
r_{A} \\
t_B^{\leftarrow}
\end{pmatrix}
= S^{-1}
\begin{pmatrix}
r^+ \\
r^-
\end{pmatrix}
\quad \text{and} \quad \begin{pmatrix}
t_{A} \\
t_B^{\rightarrow}
\end{pmatrix}
= S^{-1}
\begin{pmatrix}
t^+ \\
t^-
\end{pmatrix}\, ,
\label{B2.1}
\end{equation}
the physical amplitudes for the waveguides $A$ and $B$ shown in Eqs.~(\ref{eq.10}) are determined.

\section{A. Analysis of Transfer Probability Under Variation of $\Omega$}
\label{App.C}
\begin{figure}[h]
\includegraphics[width=0.9\columnwidth]{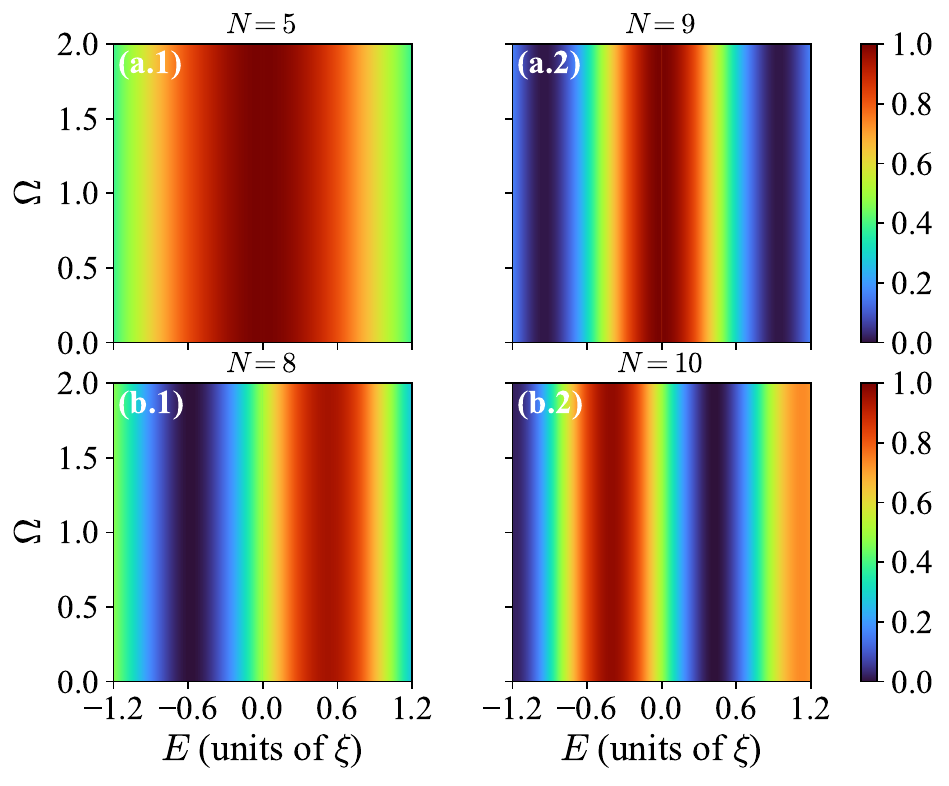}
\caption{The transfer probabilities $T_B^{\rightarrow}$ as a function of the incident energy $E$ and the coupling strength $J$ of the giant atoms. The spectra are calculated for the parameters $\omega_0 = \omega_s = \omega_e = 0$, $J = 0.01$, and $g = 1.5$ (all units of $\xi$), for (a.\,1) $N=5$, (a.\,2) $N=9$, (b.\,1) $N=8$, and (b.\,2) $N=10$.}
\label{fig8}
\end{figure}

In this appendix, we show maps confirming that the structure in Fig.~\ref{fig7}, panels~(b.1)–(b.4), remains unchanged when varying the parameter $\Omega$ with respect to the energy in waveguide~$B$. In Fig.~\ref{fig8}, panels~(a.1) and~(a.2) correspond to the energy range is approximately $E \sim [-1, 1]$ for $N = 5$ (a.1), and narrows to $E \sim [-0.5, 0.5]$ for $N = 9$ (a.2), while panels~(b.1) and~(b.2) show a shift: $E \sim [0, 1]$ for (b.1) and $E \sim [-1, 0]$ for (b.2).

\newpage
\bibliographystyle{apsrev4-1}
\bibliography{references}

\end{document}